\documentclass[aps,prd,nofootinbib,showpacs,tightenlines,superscriptaddress,preprint]{revtex4}
\usepackage{amsmath}
\usepackage{amsfonts}
\usepackage{amssymb}
\usepackage{pictex}
\newtheorem{definition}{Definition}

\usepackage[latin1]{inputenc}
\usepackage{amsmath}
\usepackage{amsfonts}
\usepackage{amssymb}
\usepackage{pictex}

\def\be{\nopagebreak[3]\begin{equation}}
\def\ee{\end{equation}}
\def\ba{\nopagebreak[3]\begin{eqnarray}}
\def\ea{\end{eqnarray}}

\def\d{{\rm d}}

\def\b{\beta}

\def\H{{\cal H}}

\def\R{\mathbb{R}}
\def\bb{{\tt b}}

\def\f{\frac}
\def\co{\sqrt{12 \pi G}}

\def\t{\tilde}
\def\l{\lambda}

\newcommand{\teta}{\rlap{\lower2ex\hbox{$\,\tilde{}$}}\eta{}}

\begin{document}

\preprint{\vbox{\baselineskip=12pt 
}}

\title{On a Continuum Limit for Loop Quantum Cosmology}

\author{Alejandro Corichi}\email{corichi@matmor.unam.mx}
\affiliation{Instituto de Matem\'aticas, Unidad Morelia,
Universidad Nacional Aut\'onoma de M\'exico, UNAM-Campus Morelia,
A. Postal 61-3, Morelia, Michoac\'an 58090, Mexico}
\affiliation{Center for Fundamental Theory, Institute for
Gravitation and the Cosmos, The Pennsylvania State University,
University Park PA 16802, USA}

\author{Tatjana Vuka\v sinac}\email{tatjana@shi.matmor.unam.mx}
\affiliation{Facultad de Ingenier\'\i a Civil, Universidad
Michoacana de San Nicolas de Hidalgo\\ Morelia, Michoac\'an 58000,
Mexico}

\author{Jos\'e Antonio Zapata}\email{zapata@matmor.unam.mx}
\affiliation{Instituto de Matem\'aticas, Unidad Morelia,
Universidad Nacional Aut\'onoma de M\'exico, UNAM-Campus Morelia,
A. Postal 61-3, Morelia, Michoac\'an 58090, Mexico}

\begin{abstract}
The use of  non-regular representations of the Heisenberg-Weyl
commutation relations has proved to be useful for studying
conceptual and technical issues in quantum gravity. Of particular
relevance is the study of Loop Quantum Cosmology (LQC), a symmetry
reduced theory that is related to Loop Quantum Gravity, and that
is based on a non-regular, polymeric representation. 
Recently, a soluble model was used by Ashtekar, Corichi and Singh to  
study the relation between Loop Quantum Cosmology and the standard  
Wheeler-DeWitt theory and in particular the passage to the limit in  
which the auxiliary parameter (interpreted as ``quantum geometry  
discreetness") is sent to zero in hope to get rid of this `regulator  
ambiguity' in the LQC dynamics. 
In this note we outline the first steps toward reformulating this
question within the program developed by the authors for studying
the continuum limit of polymeric theories, which was successfully
applied to simple systems such as a Simple Harmonic Oscillator and
the Free Particle.
\end{abstract}

 \maketitle

\section{Introduction}

Loop quantum gravity has become one of the most popular,
non-stringy, approaches to quantum gravity \cite{lqg}. 
With the aim of
understanding it better a  simpler, mechanical, model known as
{\it polymer quantum mechanics} was developed, where a system with
a finite number of degrees of freedom replaces a field theory.

The kinematical basis of polymer quantum mechanics is based on a
non-regular representation of the Heisenberg-Weyl algebra (which
turns out to be inequivalent to the Schr\"odinger's representation).
Of course, von Neumann's theorem tells us that such representation
must suffer from some kind of pathology. The pathology in this
case is the inability of polymer quantum mechanics to capture the
usual topology of the real line corresponding to the position
observables \cite{AFW}. This pathology is responsible for the non
existence of the limits suggested by a direct quantization of the
Hamiltonian (or any other observable which involves a polynomial of
the momentum). In the same fashion, within full loop quantum
gravity, the unique diffeo invariant quantization \cite{lost}
is such that the 
holonomy observables which lie at its core do not
enjoy of the continuity properties suggested by the topology of
the space manifold, and the limits suggested by the regularization
of the curvature do not exist.

A more specific motivation to better understand polymer quantum
mechanics is that it is the mathematical basis of loop quantum
cosmology (LQC) \cite{lqc}. In complete analogy with the full theory, the 
polymeric representation used in LQC is such that the 
holonomies are well defined but the connection and curvature are not 
and, furthermore, the intuitive limits that one would use
to define them though holonomies do not exist. 
The curvature is therefore defined by fixing a minimal area $\lambda^2$ when computing the holonomies, that is
motivated by the discreetness of the quantum geometry in LQG \cite{lqg}.

There is a general proposal to define effective theories, coarse
graining and continuum limits in a Wilsonian manner within the
framework of loop quantization \cite{LQasContLim}. In this context
the renormalization process modifies the continuum limit relevant
for dynamics in a way that may enhance its continuity aspects.
Recently the authors applied this proposal to a model of the
simple harmonic oscillator in polymer quantum mechanics
\cite{CVZ1}. The resulting dynamics in the continuum limit was
shown to be equivalent to that of the Schr\"odinger representation
of the simple harmonic oscillator (compare to  the original
treatment of the model \cite{AFW}). The continuum limit in this
context is different from the kinematical continuum limit because
the renormalization prescription which runs the renormalization
process demands that the Hamiltonian be physically meaningful.
This contribution targets specifically
loop quantum cosmology, so we will direct our efforts in that
direction. In particular, 
a natural question is whether such results can be generalized to LQC,
in which case one would be interested in studying
the resulting continuum
limit and its relation to the Wheeler DeWitt theory.



In this contribution we will set the stage to analyze a simplified
model of loop quantum cosmology introduced by Ashtekar, Corichi
and Singh in \cite{ACS1} known as `sLQC', from the perspective of
the Wilsonian renormalization introduced in \cite{CVZ1} to polymer
quantum mechanics. In the next section we will recall the basic
formalism of \cite{CVZ1}. Later we review the solvable model of
loop quantum cosmology of \cite{ACS1} and give a summary of what
is known about the renormalization of this model. We will close
our contribution with a list of open issues regarding the
application of our formalism to loop quantum cosmology, we will
give some remarks and some partial conclusions.

\section{Effective theories, coarse graining and continuum limit}

In this section we define the concept of effective theories and
their continuum limit. The first ingredient is the definition of a
scale to which an effective theory is associated. For a detailed
treatment of the subject presented in this section see
\cite{CVZ1}.

The role of scales will be played by regular decompositions of
$\R$ as a disjoint union of closed-open intervals of length $a_n =
\frac{a_0}{2^n}$.

\begin{definition}[Scale]
In our context, a scale $C_n$ is a decomposition of the real line
of the form $\R = \cup_{\alpha_i \in C_n}\; \alpha_i, $ where
$\alpha_i= [ L(\alpha_i) , R(\alpha_i) )$, and the vertex set is
$\{ L(\alpha_{i+1})= R(\alpha_{i})= i a_n\}_{\alpha_i \in C_n}$.
\end{definition}

To every scale $C_n$ we associate a space of {\em states at scale}
$C_n$, $Cyl(n)$,
with a basis $\{ e_{\alpha} \}_{\alpha \in C_n}$ labeled by the
cells of $C_n$. The inner product is given by $ ( e_{\alpha_i},
e_{\alpha_j} )_{C_n} = \delta_{ij}
$. The completion of $Cyl(n)$ is the Hilbert space at the scale
$C_n$,  $\H_n$.
This inner product is inherited from the corresponding polymer
Hilbert space, see \cite{CVZ1}.

We will also work with the dual space $\H_n^\star$. Since the
space of states at any given scale is a Hilbert space, its dual is
isomorphic to it, but for us the dual space will be of special
interest. Notice that its elements have a natural
$\sim_n$-preserving action on $Cyl(\R)_x$ (the vector space of
cylindric functions on the real line). This action is particularly
simple to see for the elements of the dual basis $\{
\omega_{\alpha} \}_{\alpha \in C_n}$; for them we have
$\omega_{\alpha} (\delta_{x_0})= \chi_{\alpha} (x_0)$ where
$\delta_{x_0}(x)=0$ if $x\ne x_0$ and  $\delta_{x_0}(x)=1$ if
$x=x_0$, and $\chi_{\alpha}$ is the characteristic function of the
set $\alpha \subset \R$. Thus, we write $ \H_n^\star \subset
Cyl(n)^\star $ where by $Cyl(n)^\star$ we mean the
$\sim_n$-preserving subspace of $Cyl(\R)_x^\star$.

In order to define the continuum limit of the effective theories
we have to relate different scales by mapping between the
corresponding Hilbert spaces and its duals.
\begin{definition}[Coarse graining]
Given two scales we write $C_m \leq C_n$ and say that $C_m$ is a
coarse graining of  $C_n$ (or $C_n$ is a refinement of  $C_m$) if
any interval $\alpha_i \in C_m$ is a finite union of intervals of
$C_n$.

Our coarse graining maps work by decimation. If we have two scales
related by refinement $C_m\leq C_n$ our decimation map will be
defined to be the injective isometry $d:\H_m \to \H_n$
characterized by $ d(e_{\alpha_i}) = e_{\beta_j} \quad \iff
L(\alpha_i) = L(\beta_j) . $ It is important to notice that if
$C_m\leq C_n \leq C_o$ the corresponding $d$-triangle diagram
commutes.\footnote{It is called coarse graining because these mappings
induce a `coarse graining' of the measure and the observables.}
\end{definition}

On the other hand, $d^\star:\H_n^\star \to \H_m^\star$ sends part
of the elements of the dual basis to zero while keeping the
information of the rest: $d^\star (\omega_{\beta_j})=
\omega_{\alpha_i}$ if $j=i2^{n-m}$, in the opposite case $d^\star
(\omega_{\beta_j})= 0$.
%


At any given scale we have an effective theory and we can
calculate the expected values of every observable at this scale.
For an observable $\hat{O}$ we consider its expectation value on
normalized states $\kappa_{C_n}^2 ( \psi , o_n   \psi )_{C_n} =
o_n(\psi)$, $ o_n : \H_n \to \R . $ The normalization factors in
the inner product $\kappa_{C_n}^2 \in \R^+$ have to be adjusted in
such a way that, at least in the continuum limit, the observables
of different scales are pasted
correctly by the decimation maps.%
\footnote{In standard introductions the continuum limit in the
renormalization group framework includes a wave function
renormalization; here we choose the equivalent action of
renormalizing the inner product instead. We choose to absorb the
normalization factors in the inner product to find a non trivial
action of the completely renormalized observable $\hat{O}$ in
$\H_{{\rm
poly},x}$. } 

At a given scale $C_m$ we can ``include the effects of more
microscopic degrees of freedom" by using our decimation map. When
$C_m \leq C_n$ we define
\be o_{m(n)} := d_{m,n}^\star o_n .\label{cond-a} \ee
That is, $o_{m(n)}(e_{\alpha_i}) := o_n( d_{m,n}e_{\alpha_i})$. If
these microscopically corrected observables converge, their limit
will be called a completely renormalized observable at the given
scale $o_m^{\rm ren} : \H_m \to \R$,
\be o_m^{\rm ren} := \lim_{C_n \to \R} o_{m(n)} .\label{cond-0}
\ee
By construction, when the completely renormalized observables
exist, they are compatible with each other in the sense that $
d_{m,n}^\star o_n^{\rm ren} = o_m^{\rm ren} . $ A collection of
compatible observables defines in itself a continuum limit
observable.

A natural choice of normalization factors that leads to
convergence in (\ref{cond-0}) for the case of the simple harmonic
oscillator (SHO) \cite{CVZ1} is $\kappa_{C_n}^2 = 2^n$ which means
that the renormalized inner product in $\H_n^\star$ is $
(\omega_{\alpha_i} , \omega_{\alpha_j})_{C_n}^{\rm ren} =
\frac{1}{2^n}
\delta_{ij} $
\footnote{ In the case of LQC the symmetric gravity plus matter
constrained system is parametrized as an unconstrained system with
one degree of freedom evolving in the relative clock of the
matter. That parametrized system is then represented in a Hilbert
space at ``scale'' $n$ which is similar to the $\H_n$ defined
above, but whose inner product is not $\delta_{ij}$. A
renormalized inner product for the dual Hilbert space would have
to be defined following our procedure. }. The Hilbert space of
covectors together with such inner product will be called
$\H_n^{\star{\rm ren}}$. A sequence of covectors $\{
\Psi_{C_n}^{\rm ren} \in \H_n^{\star{\rm ren}} \}$ is called
compatible if $d^\star \Psi_{C_n}^{\rm ren} = \Psi_{C_m}^{\rm
ren}$ for each pair with $m \leq n$. These compatible sequences
form the Hilbert space
$\stackrel{\longleftarrow}{\H}_{\R}^{\star{\rm ren}}$ and the
inner product in this space is $( \{ \Psi_{C_n} \} , \{
\Phi_{C_n}\} )^{\rm ren}_{\R} := \lim_{C_n \to \R} ( \Psi_{C_n} , 
\Phi_{C_n} )^{\rm ren}_{C_n} $. Observables in this continuum
limit $o_{\R}^{\rm ren} : \stackrel
{\longleftarrow}{\H}_{\R}^{\star{\rm ren}} \to \R$ are defined by
\be o_{\R}^{\rm ren} (\{ \Psi_{C_n} \}) := \lim_{C_n \to \R} o_n
((  \Psi_{C_n} , \cdot )^{\rm ren}_{C_n}) \label{w.o.cut-off} .
\ee
The continuum limit of the effective theories exists if the above
limit exists for enough observables of physical interest.

A physical Hilbert space can be defined when one considers the
degeneracy of the inner product and the observables that we just
defined. In the case of SHO it is shown that the corresponding
physical Hilbert space is unitarily isomorphic to $L^2(\R ,\d x)$,
the usual Hilbert space of the Schr\"odinger theory\cite{CVZ1}.
\section{Loop Quantum Cosmology}

In this section we shall follow \cite{ACS1} closely in order to
define the model of interest. The gravitational phase space
variables in the homogeneous, isotropic, spatially flat sector of
general relativity
can be expressed as, \be A_a^i=c\,V_0^{-(1/3)}\,{}^{\rm
o}\omega^i_a \qquad{\rm and}\qquad E^a_i=p\,
V_0^{-(2/3)}\,\sqrt{q_0}\,{}^{\rm o}e_i^a \ee where $V_0$ is the
volume as given by the fiducial metric ${}^{\rm o}q_{ab}$ of an
auxiliary cell that is used to make the integrals in the
variational principle (and canonical formulation) of homogeneous
models finite and ${}^{\rm o}\omega^i_a$ is the dual basis to a
fiducial triad ${}^{\rm o}e_i^a$, compatible with ${}^{\rm
o}q_{ab}$. In terms of geometrodynamical variables $|p|$ is
proportional to $a^2$, where $a$ is the scale factor of the FRW
metric, and $c$ is proportional to $\dot{a}$ (which determines the
extrinsic curvature).

The fundamental Poisson bracket for the gravity variables is given
by, $ \{c,p\}=\frac{8\pi G\gamma}{3} $, where $\gamma$ is the
Barbero-Immirzi (BI) parameter that parametrizes an ambiguity in
the loop quantization.

Defining the new coordinates, \be {\tt
b}:=\frac{c}{|p|^{1/2}}\qquad{\rm and} \qquad {V}:=p^{3/2} \ee $V$
is the physical volume of the auxiliary cell. The new coordinate
${\tt b}$ is then equal to ${\tt b}=\frac{c}{a}$. On the
constraint surface, that is, on classical solution to the
equations of motion it becomes ${\tt
b}=\frac{\gamma\dot{a}}{a}=\gamma\,H_{\rm Hubble}$. The Poisson
bracket now becomes, $ \{{\tt b},{V}\}=4\pi G\gamma $. In these
new coordinates, the Hamiltonian constraint of the gravitational
part coupled to a massless scalar field $\phi$ is: \be {\cal C}:=
-\frac{6}{\gamma^2}\,V\,{\tt b}^2+8\pi G\,\frac{p^2_\phi}{V}=0 \ee
with $p_\phi$ the momentum conjugate to $\phi$. One has to define
a corresponding operator $\hat{\cal C}$. But on the polymeric
Hilbert space $\hat{{\tt b}}$ is not well defined. In order to
`regulate it', the natural basis to consider is: $
|\nu_n\rangle=|\lambda\,n\rangle$, with
$\hat{V}\,|\nu_n\rangle=2\pi\ell^2_{\rm p}\,\hat{\nu}|\nu_n\rangle
= 2\pi\ell^2_{\rm p}\,\nu_n\,|\nu_n\rangle $ \noindent That is,
the lattice has uniform spacing in the coordinate $\nu$ with
spacing given by $\lambda$, the parameter that dictates the scale
as understood in the previous section. Note that it is an
eigenbasis of the volume operator. A wave function $\t\Psi(\nu)$
has only support on the discrete set $\nu_n=\lambda\,n$.
Intuitively, in the regularization one is going to replace $\bb$
for: $\hat{\bb} \longrightarrow \frac{\widehat{\sin(\lambda
\bb)}}{\lambda}$ (see \cite{ACS1} for the precise construction).
The constraint is
\be \label{hc4}
\partial_\phi^2\, \tilde{\Psi}(\nu,\phi) = 3\pi G\, |\nu|\,
\f{\sin\lambda\b}{\lambda}\, |\nu|\, \f{\sin\lambda\b}{\lambda}\,
\tilde{\Psi}(\nu,\phi) \ee which takes the form \cite{ACS1}:
\be \label{hc5} \partial_\phi^2 \,\tilde{\Psi} (\nu, \phi)
=\f{3\pi G}{4\lambda^2}\, \nu \left[\, (\nu+2\lambda)
\t\Psi(\nu+4\lambda) - 4\nu \t\Psi(\nu) + (\nu -2\lambda)
\t\Psi(\nu-4\lambda)\, \right] \ee With the physical inner product
given by
$$(\t\Psi,\t\Phi)_{\rm phy}=
\sum_{n\in \mathbb{Z}-0} \frac{1}{|4n\l|}\;\overline{\t\Psi}(\nu_n)\,\t\Phi(\nu_{n})
\; .
$$
The physical inner product (and Hilbert space) in this section
refers to those objects that are needed in order to make physical
predictions from the solutions of the quantum constraints for
constrained systems (as is the case here). This has to be
contrasted with the physical Hilbert space defined as an
appropriate limit in \cite{CVZ1}.

Let us now take functions of $\Psi({\tt b},\phi)$. In this case
the quantum constraint of sLQC (after a rescaling
$\chi=\Psi/\nu$), is given by,
\be \f{\partial^2}{\partial \phi^2}\cdot\chi(\bb,\phi)=12\pi G\,
\left(\frac{\sin(\lambda\bb)}{\lambda}\;\frac{\partial}{\partial\,
\bb} \right)^2 \cdot \chi(\bb,\phi)\label{const1} \ee
\noindent with $\bb\in (0,\pi/\lambda)$ and
$\chi(\bb,\phi)=-\chi(\pi/2\lambda-\bb,\phi)$. Let us note that
since the variable $\nu$ is discrete, its canonically conjugate
variable $\bb$ is compactified \cite{CVZ2}. If we define a new $x$
coordinate as: $x = \f{1}{\co} \, \ln \left(\tan\left(\f{\lambda
\bb}{2}\right)\right)$, then the basic constraint equation
(\ref{const1}) translates to $\partial_\phi^2 \, \chi(x,\phi) =
\partial_x^2 \, \chi(x,\phi)$. A general solution  $\chi(x,\phi)$
to the previous equation can be decomposed in to left moving and
right moving components: $\chi = F (\phi + x) - {F(\phi - x)}$
that furthermore, satisfy the required symmetry.

In SLQC the expectation value for $\hat{V}$, is given by:
\be \langle\hat{V}|_\phi\rangle=V_+\,e^{-\sqrt{12\pi\,G}\,\phi}
+V_-\,e^{\sqrt{12\pi\,G}\,\phi} \ee There is a minimum for the
expectation value at the bounce time $\phi_b^V$ given by, \be
\phi_b^V:=\frac{1}{2\,\sqrt{12\pi\,G}}\ln\left(\frac{V_+}{V_-}\right)
\ee On the other hand, the WDW equation in the $\bb$
representation reads, \be
\partial_{\phi}^2\cdot\chi(\bb,\phi)=12\pi G[\bb\,\partial_{\bb}]^2\cdot
\chi(\bb,\phi) \ee \noindent Introduce $ y:=\f{1}{\sqrt{12\pi
G}}\ln(\bb/2\bb_0) $ , from which the equation becomes, \be
\partial_\phi^2\;\chi(y,\phi)=\partial_y^2\;\chi(y,\phi)
\ee The same equation as in LQC! What is the main difference then?
Of course, the representation of the Dirac observables
$\hat{V}_\phi$: \be {\rm WDW:}\quad \hat{\nu}:=\exp({\co\,
y})\,\partial_y \qquad ;\qquad {\rm LQC:}\quad
\hat{\nu}:=\cosh(\co\, x)\,\partial_x \ee Therefore, in the WDW
theory \be
\langle\hat{V}\rangle=\underline{V}_0\,e^{\pm(\co\;\phi)} \ee with
the sign depending on the choice of branch. For a given $\lambda$
we can have a WDW description in terms of the Klein Gordon
equation where the choice of the parameter $\bb_0$ is rather
convenient: set $\bb_0=2/\l$. In this case, if we fix a state
$F(x_+)$ for $\phi_0=0$ given, that corresponds to an initial
state with support on small values of $x\ll -1$, we can expect it to
have a small density (by fixing $p_\phi$ and having large volume).
For that class of states, we shall compare the WDW and sLQC. The
Hilbert space is the same, so the same wave function (solution to
KG) is a physical state on both theories, and the expectation
value of the volume can be arbitrary close for both theories. The
difference is given by \cite{ACS1}, \be \langle \hat{V}|_\phi
\rangle_{\rm wdw} - \langle \hat{V}|_\phi \rangle_\lambda =
V_-\,e^{-\co\,\phi} \ee with $V_-\ll 1$ the  initial difference,
that goes to zero as $\lambda \to 0$. Note that this prescription
is such that $[\langle \hat{V}|_\phi \rangle_{\rm wdw} - \langle
\hat{V}|_\phi \rangle_\lambda]\to 0$ when $\phi\to\infty$, but
that the difference grows unboundedly for the other sigh of $\phi$
(in the direction of the big bang).

This assignment that relates the `discrete theory at scale
$\lambda$' with WDW, the putative continuum theory, can also be
used for defining a 'coarse graining map' between scales
$\lambda$ and $\lambda'=\alpha\lambda$ \cite{ACS1}: Given a state
$\chi^\lambda(x,\phi)$ define the new state
$\chi^{\lambda'}(x,\phi)=\chi^\lambda(x+\mu,\phi)$ with
$\mu=-\f{1}{\co}\ln\alpha$. This assignment ensures that the
difference in  volume at that time is very small: $ \Delta
(\langle \hat{V}\rangle)=(1-\alpha)\,V_-\ll 1 $, since $V_- \ll 1$ and
$\alpha<1$. Again, under time evolution, the difference will grow
unboundedly as $\phi\to -\infty$: \be \Delta (\langle
\hat{V}\rangle)|_\phi=(1-\alpha)\,V_-\,e^{-\co\, \phi} \ee Thus,
with this prescription the time evolution is such that the
physical observables become closer in the macroscopic domain
(where the universe is large), and only depart when approaching
the Planck domain to grow unboundedly on the pre big bang era
\cite{ACS1}. It was also shown that there is no `coarse graining map'
for which the observable $\hat{p}_\phi$ (and any
power of it) and the volume $\hat{V}|_\phi$ can preserve their
expectation values throughout the `renormalization flow', {\it for all times},
preserving desired features in the large volume, macroscopic
regime \cite{ACS1}. From this perspective one is lead to conclude
that, as the scale $\l$ is refined, there is no limiting theory.
Therefore,  it is concluded that there is no continuum 
limit and thus sLQC is, in a precise sense, fundamentally 
discrete \cite{ACS1}. 

What we still need to understand is how this `coarse graining map'
\cite{ACS1}
can be translated to the representation $\Psi(\nu)$
where the notion of scales and `lattice refinement' is natural. As
a second step, one needs to study the properties of the limiting
Hilbert spaces and the action of the Hamiltonian of the theory
that generates time evolution in $\phi$. We shall leave these
matters for further investigation.

\section{Discussion}

Ashtekar, Corichi and Singh introduced a solvable model
\cite{ACS1} in order to study several aspects of loop quantum
cosmology including its relation to the more traditional Wheeler
DeWitt theory, a putative candidate for a continuum limit of LQC.
Given that the model is completely solvable, it is a rich arena to
study many aspects of loop quantum cosmology, polymer quantum
mechanics and in general to study the physics of models based on
non-regular representations of the Heisenberg-Weyl algebra. Their
study of the continuum limit of the sLQC model, understood as
approaching the limit in which the discrete nature of the quantum
geometry is expected to go to the continuum, involved considering
the behavior of the volume observable and its evolution (according
to the internal clock defined by the scalar matter, and fixing the
other observable of the theory). In the model there is not one
single evolution operator, but one for each ``auxiliary scale''.
The results of \cite{ACS1} show that, as  the ``auxiliary scale''
is refined, the evolution of the volume observable does not
converge to a hypothetical continuum theory (and therefore neither
to the Wheeler DeWitt theory).

The work in \cite{ACS1} teaches us some aspects of the dynamics of
the solvable loop quantum cosmology model, and about a specific
`coarse graining map' that is naturally suited to the
formulation of the model as a solvable system. However, a detailed
study of the  model according to the framework of effective
theories, coarse graining and continuum limit for polymer quantum
mechanics as understood in \cite{CVZ1}, remains to be undertaken. In
particular, the physical Hilbert space of the continuum as defined
in \cite{CVZ1} has not been explicitly constructed and the
behavior of the observables and the dynamics thereon has yet to be
studied. Work along this line is in progress.

A second and important question pertains to the conclusions one
might draw from either result coming out of such investigations.
In particular there might be very different expectations depending
on some personal bias:
For instance, some people may not find the notion of ``scale'' as
natural in this case as it was in the case of polymer quantum
mechanics.
Therefore there may be several conclusions one may reach from the
non existence of the limit studied in \cite{ACS1}. Here we show
two lines of thinking: A) The system is fundamentally discrete and
there is a ``scale'' $\lambda$ such that the dynamics encoded in
the constraint of sLQC at that value of $\lambda$ is the
fundamental dynamics. The constraints which use other values of
the parameter are just incorrect; they are not effective theories
that describe the system approximately. B) A ``removal of the
cut-off'' by the limit studied in \cite{ACS1} is not the correct
way to proceed. Instead, one may try to follow the lines of
\cite{CVZ1} in which at each scale one has an effective theory and
at each such scale one can import corrections from smaller scale
by a coarse graining procedure. The resulting mathematical
structure of the limits taken in such an approach is
different from those of  \cite{ACS1} and needs to be studied.

If the continuum limit suggested by \cite{CVZ1} does not exist,
the coarse grained effective dynamics does not become better
approximations to ``the true system'' as the scale is refined and
thus either the effective dynamics at a given scale or the coarse
graining procedure are wrong. In this case our procedure of
non-perturbative renormalization would fail to define a theory of
sLQC in the continuum. To understand which of these scenarios is
realized is a pressing matter. 

\begin{acknowledgments}
We thank A. Ashtekar and P. Singh for discussions. This work was
partially supported by CONACyT U47857-F and NSF PHY04-56913 grants
and by the Eberly Research Funds of Penn State.
\end{acknowledgments}

\end{document}